# Propagation Model for SSC attacks: Why SBOM (tools) don't tell the whole truth

Ljubica Grgic*[1]; Lazar Maksimovic[1]; Pavel Laskov[1]

*Corresponding Author
[1] University of Liechtenstein, Vaduz, Liechtenstein
ljubica.grgic@uni.li; lazar.maksimovic@uni.li;
pavel.laskov@uni.li

**Abstract.** Ensuring security of software supply chains (SSC) is indispensable in today's world of modern software practices. SBOM (tools) have been introduced as relevant building blocks to ensure the transparency of SSCs. However they have serious limitations in practice as their vulnerability detection and interpretation capacity is not sufficient to explain exploitability effects that can propagate through the whole chain. To address this gap, we propose a propagation-centred approach to SSC security and introduce a four-stage propagation model. We empirically evaluate four open-source SBOM tools against each stage using three projects and Log4j vulnerability as our test case. Our results show that current SBOM tools systematically support only Stage 1 (Structural Exposure) and Stage 2 (Vulnerability Class Presence) while Stage 3 (Code Reachability) and Stage 4 (Taint Path Analysis) require capabilities absent from the SBOM ecosystem. We argue that putting propagation effects at the centre of SSC security research is essential to prevent cyber risk evolving into systemic risks. Our research findings contribute to a future research and design of modern SSC security tools.

**Keywords:** Software Supply Chain Security (SSCS), SBOM, Propagation.

## 1 Introduction

What have NotPetya, SolarWinds and Log4Shell (initial outbreaks in 2017, 2020 and 2021) in common? Next to their somewhat cryptic names, they belong to what is commonly defined as an *attack* on a Software Supply Chain (SSC). More important in the context of this paper, their common characteristics is having a similar attack surface approach. Specifically, a SSC attack occurs not by targeting the end-user organization directly but instead targeting e.g. its upstream dependencies, infrastructure, update mechanism or third-party components. What it separates from traditional cyberattacks (traditional in the sense that the target is one single organization) is their mechanism of exploitation. This means, SSC attacks succeed because they exploit the "trust" relationships prevalent in nowadays software development practice, where applications incorporate multiple open-source libraries, third-party components etc. as dependencies. This then gives a whole new dimension to the cyber risk. Consequences become systemic because of the propagation effect. As seen in real life, propagation effects happened simultaneously (and even undetected for some time) with the SolarWinds compromise in 2020 and the Log4Shell vulnerability in 2021. Log4Shell even became a reference point for how a single vulnerable component could propagate through the entire digital ecosystem [1].

SSC attacks have made one thing very apparent: many organisations have little and/or insufficient knowledge of the dependencies and its behaviour within their software supply chain [2]. What is the consequence resulting from this lack of knowledge? Due to the fact that any vulnerable software component may propagate through the SSC and as such increase the risk for the whole software system and beyond [3] the challenge is intensified to maintain and ensure security of the SSC.

It then does not come as a surprise, that the European Union Agency for Cybersecurity's (ENISA) has defined software supply chain (SSC) attacks in their Foresight 2030 report [4] as one of the most fundamental threats organizations could face. The report shows the industry sentiment where the compromise of software supply chains is predicted to remain a top cybersecurity threat by 2030 (and even listed as the number one (out of ten) threat). The rationale behind is clear: vulnerabilities in SSCs cause more serious security threats than independent software systems because of their complex interdependency and dependency relations. The lack of visibility into relations, composition and functionality of these systems contributes to an increase in cyber security risks [5]. However, ensuring transparency and visibility is becoming challenging as at the same time the reliance on third-party components and dependencies is increasing.

### 1.1 The importance of understanding Propagation in the context of SSCS

In the context of cyber risk and software supply chain security the above mentioned dependency relationships often lead to propagation effects. Propagation effects are defined as the spread of a vulnerability across interconnected systems and/or software components. Meaning, it spreads beyond the initially compromised entry point and causes therefore cascading effects within the software supply chain. One very concrete real-world example proving the high relevancy in understanding the propagation behaviour of SSC attacks are the long-term effects vulnerability exploitations. As shown by [6] and [7] in a longitudinal measurement study of Log4Shell (period from 2021-2025) their results demonstrate that Log4Shell remains active years beyond its initial disclosure period. Active in the context of their study means that exploitation still persists and they showed that it had become more extensive and sophisticated in the process. Even if the remediation path is well-understood and the vulnerability of Log4Shell is widely known the vulnerable versions still continue to circulate. It is fair to conclude that in this case propagation still continues.

To summarize, cyber risk can transform into a systemic risk that propagates through whole ecosystems. Hence, we conclude that there is a demand for new understandings in ensuring software security that include analysis on propagation effects. Before we present our approach in this regard, it is necessary to shed a light on the current status quo on existing tools that help securing the SSC (so far).

**Existing tools used to secure SSC.** Log4Shell (also) started the discussion of a new era of software supply chain security such as SBOM mandates, stricter oversight in Europe through Network and Information Systems Directive (NIS2) and the EU Cyber Resilience Act (CRA), and an increased investments in open source security [1]. The question on how to secure a SSC has prominently been addressed through the concept of Software Bill of Materials (SBOM) [5]. SBOM have been introduced as relevant building blocks to ensure the transparency of software supply chains [8]. This is reached by SBOMs representing component relationships within a software product and, more significant, supporting software composition analysis (SCA) by linking components to known vulnerabilities [9]. As stated also in [2] SBOM has recently emerged as a key concept to enable principled engineering of software supply chains, listing the necessity of accurate SBOMs in order to track vulnerabilities, detect them and as a result ensure SSC integrity.

Now, referencing back to the very first sentence of our paper where we started by mentioning some of the most prominent SSC attacks it is a this point legitimate to ask “Why is it then, that SBOMs fail in securing the SSC”? It is clear that somewhere between the theoretic idea and practical application expectations are not met. This argument can be strengthened by the practical example of the Log4j attack. As this component was extensively used by a large number of open-source and proprietary projects, it was tiresome and costly to identify all impacted projects. The analysis performed by [2] showed that if all these Java projects had published a SBOM, it would have facilitated the precise identification and remediation of the vulnerable applications. But even organizations with complete SBOMs still struggled because the vulnerability existed in a library that was ubiquitous and created a manifold of dependencies that could not be easily updated [10]. As a matter of fact this is visible even four years later after the initial outbreak. In 2025 nearly 300 million of total Log4j downloads happened and off those, about 13% represent still vulnerable versions [1]. Clearly the SSC is still not secure as it could be (as safe versions are indeed available). With this obtained information, we ask ourselves what could be the reason, that SBOM is not as effective in practice as it is supposed to be?

Looking at the Figure 1 below that represents a vulnerability report a SBOM tool created – how can it help us finding vulnerabilities? Does this figure give an indication on why SBOMs fail in securing the SSC?

```
"Severity": "CRITICAL",
"CweIDs": [
  "CWE-20",
  "CWE-400",
  "CWE-502",
  "CWE-917"
],
"VendorSeverity": {
  "amazon": 4,
  "ghsa": 4,
  "nvd": 4,
  "redhat": 4,
  "ubuntu": 3
},
```

```
"CVSS": {
  "ghsa": {
    "V3Vector": "CVSS:3.1/AV:N/AC:L/PR:N/UI:N/S:C/C:H/I:H/A:H/E:H",
    "V3Score": 10
  },
  "nvd": {
    "V2Vector": "AV:N/AC:M/Au:N/C:C/I:C/A:C",
    "V3Vector": "CVSS:3.1/AV:N/AC:L/PR:N/UI:N/S:C/C:H/I:H/A:H",
    "V2Score": 9.3,
    "V3Score": 10
  },
  "redhat": {
    "V3Vector": "CVSS:3.1/AV:N/AC:L/PR:N/UI:N/S:U/C:H/I:H/A:H",
    "V3Score": 9.8
  }
},
```

**Fig. 1.** Apache Solr 8.11.0 Trivy Scan Showing Log4j Log4Shell Vulnerability

It is fair to conclude that the detection capability of SBOM tools don't go beyond the binary presence/absence of a CVE flag (which is the current state of practice) as highlighted in Figure 1. It says that affected Log4j versions can execute attacker-controlled code when the application logs a specially crafted string, usually through Jndiookup behaviour. The severity is marked CRITICAL, with CVSS (Common Vulnerability Scoring System) scores up to 10, meaning it is extremely serious and can allow full compromise of a vulnerable server.

While SBOM is a tool used with the goal to secure a SSC, research [11] has shown that its practical utility is undermined by inaccuracies in both its generation and its application in vulnerability scanning. Empirical evidence [11] is provided that reachability [12] analysis is a fundamental requirement currently missing in vulnerability scanning tools.

Why is this relevant for SSC security? Following the introduced argument of propagation effects, code reachability analysis can provide insights into those dynamics. Reason being, if the code is not reachable, the vulnerability cannot propagate to an exploit, hence the cyber risk is predictable and therefore controllable. Put differently, it is relevant to know if the "path" to the code is open (or not). This would be the desired state of knowledge to have (to design an effective and proactive defence strategy around the software supply chain). Currently SBOM does provide visibility but does not provide control.

To conclude with a generalized statement, the factual circumstance in practice is that existing vulnerability management tools are struggling to keep up with the rapid evolution of the software it aims to protect. Sonatype's security research [7] revealed that it is the entire system of tools and frameworks that allows open source vulnerabilities to exist. What is more, they claim that despite major investment in scanning tools, disclosure pipelines, and security automation, organizations continue to operate with blind spots. This again undermines the necessity to think about new solutions.

### 1.2 Problem Statement and Our Contribution

**Problem Statement.** SBOM tools can identify the presence of a vulnerable dependency, but cannot determine whether the vulnerable method is reachable or whether attacker-controlled input can reach a vulnerable logging sink. Hence they do not give information on propagation effects. This is a problem because in order to control exploitability it is necessary to know to where the vulnerability has propagated already.
**Contribution.** To address the above mentioned limitations of SBOM tools in SSC context, we propose one very concrete approach on how to make practical sense of dependencies in SSCs with the goal to increase security. Making practical sense of dependencies means to gather and grasp information relevant to depict potential propagation effects. With this in mind, one of the contributions of our paper here is to introduce a four-staged propagation model that could serve as a foundation for further research on modern tools that do analyse information related to propagation effects (such as reachability of code etc.). This four-stage model provides a structured and reproducible approach. In addition, we will contribute with an experiment in evaluating four open-source SBOM tools against each of the four stages. As a result, we will show that no evaluated SBOM tool provides detection capability beyond a certain stage that would be required to provide information on potential propagation effects of exploitability. Finally, we do show in our experiment that there are tools with whom it is possible to detect exactly those two elements of propagation.

**Definition of Propagation.** For the experiment as outlined further below propagation will be distinguished and analysed according the following two definitions (which are mutually inclusive):

1. **Structural propagation or dependency chain propagation:** means, the way a vulnerability embedded in one component (which will be our example of the library like Log4j) is inherited by every software product that depends on it, directly or indirectly.
2. **Dynamic propagation: Exploit propagation or cascading effect**: means how successful the exploitation was in enabling a lateral movement that lead to cascading compromise of further (eco)systems. The (mutually exclusive) questions to be answered here are the following: is the vulnerability reachable? If yes, what about the effectiveness of the input - is it enough to practically exploit the reachable vulnerability?

## 2 Related Work

### 2.1 Relevance of understanding SSCs risk to manage cyber risk

Following the magnificent (in effect and damage) supply chain attacks the importance in analysing supply chain networks in the context of cyber security has been recognized [14] as relevant for providing significant insights in understanding and even to some extent predicting cyber risk for the whole organisation and respective (eco)system in scope. For example, the study's [13] main quantitative results shows that supply chain network features add significant detection power to predicting enterprise cyber risk. They state also further research has to be done to better understand the causal mechanisms and the specific vectors of attack, and what risk indicators best capture them. Their limitations, as the authors themselves conclude, that these results don't go further such as to design and propose a novel framework on managing cybersecurity risks with more effective visibility, assessments, insights, etc. Finally, they conclude that current guidelines lack to certain extent the dynamic in capturing the propagation effects of a cyber risk in a supply chain.

### 2.2 Current research status on limitations of SBOM (tools)

Looking first on the practitioners side on SBOM and the limitations they see, the main critics arise around the fact that SBOM was designed as a static inventory and vulnerability matching tool. Therefore, practitioners generate SBOMs to satisfy compliance requirements, rather than to ensure having visibility into their supply chain risk [10] as SBOM tells what is around but it doesn't tell anything about the security status of the components. Also, it does not give any information on how those components arrived.

Currently the conversation has started to shift to SBOM consumption as opposed to a mere SBOM generation. In the context of our example on Log4j a (future) SBOM should not only answer "do I have Log4j?" but "can Log4j actually be exploited in my environment?". As stated by [10] it has to include information on reachability analysis, network exposure etc. Our experiment further below will show at which stage todays SBOM tools stop and what is needed to consider reachability analysis.

Some researcher already performed systematic literature reviews (SLR) on SBOM. In the following section we will therefore highlight the main arguments we consider relevant for our paper here. In their SLR the authors [14] tried to find an answer (among others) for their research question “What barriers impact the use of SBOMs for SSC security?” their conclusion to that is that more work is needed around exploitability (in their argument they propose an automation). More important for our paper here, they found that dynamic risk modelling is missing and argue therefore to enhance current SBOM tools with “runtime telemetry approaches”[15].

Talking now about the element of accuracy in the context of SBOM and SSCS it is really foundational (and we argue also a prerequisite) for an effective usage of the tools in practice. In this context the authors [11] highlighted two main missing aspects as proposal for further future work and research: 1) the integrity of the input (ground truth) has to be ensured by using lock files (instead of project files, as commonly used by practitioners) and 2) next generation of SBOM tools (they focus especially on vulnerability scanners) have to incorporate reachability analysis. However the interesting argument for our paper here they provide is the following: While increasing the accuracy by using lock files is a crucial first step, their study [11] reveals that even then (meaning even if SBOM tools become more accurate) the overwhelming 97.5% false positive rate couldn’t be reduced. Clearly, something is missing here.

Research was also conducted in analysing the quality of SBOM tools and the respective tools ecosystem. As stated [16] at least two challenges exist with SBOM adoption and have not yet been addressed: 1) there is an expectation of purpose for its usage and it is often stated very generically (e.g. to ensure transparency, component management). This brings a practical problem to the usage, because without knowing the purpose for creating an SBOM, users may simply fail to provide the relevant information. And the second point mentioned in [16] refers to the necessity of having dependable automated tools where a validation of SBOMs (and tools) and an assessment on their quality is mandatory.

An empirical evaluation of multiple SBOM generation tools was conducted [17] whereby the authors assessed their effectiveness and operational efficiency. Their results underscore the importance of tailoring tool selection to the characteristics of the target infrastructure. Also they highlight a notable deficiency in the standardization of methodologies for evaluating the performance and accuracy of SBOM generation tools.

Part of our literature review included also looking at related work talking more on a governance level. That is relevant to include because the whole concept of SBOM was initiated to enhance a better governance of the software supply chain and to manage third-party risk. Focusing on the effectiveness of frameworks for software supply chain security [18] the research wanted to find out what tasks actually mitigate attack techniques. Reason for that lies in practical nature - to give users instructions on how to address specific threats and close (potential) mitigation gaps. The analysis included prominent attacks such as SolarWinds, Log4j, or XZ Utils. One of their main results was that at least three [19] mitigation tasks were missing from all analysed (ten) frameworks. These findings demonstrate that the current generation of SBOM tools is fundamentally inadequate for practical use and in capturing the very specific nature of cyber risk in a software supply chain. Along with that, we found only one explicit research paper [20] proposing an advanced SBOM tool (calling it UniSBOM) that includes SBOM generation, analysis and visualisation in one. The authors' main motivation was to enhance the security accountability of networked systems (IoT) with their newly introduced UniBOM. They address the limitations of existing SBOM tools by enabling a more fine-grained approach to SBOM creation and seamless vulnerability analysis. Effectiveness of their tool is demonstrated by its superior detection capability. While UniBOM enhances visibility and accountability in dependable systems it does not address the propagation effects (such as reachability) we highlighted already as being fundamental.

We can summarize this section on related work by saying that SBOMs just give a list of ingredients at a certain point in time without going further in saying if there is a possibility that those components could introduce a risk into the environment and neither do they tell, how the spreading effect of this risk can be. This underlines the necessity to address it from a different perspective beyond solely increasing the visibility as seen in the related work we read. The next section introduces different elements from related work we consider relevant to contextualize our later proposed propagation model.

### 2.3 The Reachability Problem, Attack Path and Propagation Effect

As mentioned above, the (still) unresolved reachability problem is very crucial in understanding the cyber risk within software supply chains as it is the point where actually propagation can be depicted to be predicted. This then could help in developing new defence strategies. While it is agreed upon [21] that the challenging problem still remains in determining the exact conditions under which a vulnerability in a third-party component may affect a dependent program a good approximation can be obtained by considering the reachability of the vulnerable code from the parent program. This argument has been further detailed out by [22] where the authors confirm that current SBOM techniques are not enough to track dependencies from the source code up to the service-oriented level, nor from the design to the operation phases. As we already claimed, the authors strengthen our argument and confirm that these limitations prevent discovering the propagation of the vulnerability. This then leads to the reality that neither a detailed operational impact analysis in a System-of-Systems (SoS) context is possible, nor a lateral movement analysis [22]. To the best of our knowledge, we have found this research paper to be one of the only ones that actually propose a solution to the problem of tracking propagation. They propose three dependency graphs in order to be able to overcome the previously described challenges that prevent analysing the propagation effect. In a similar direction but with a different approach they [23] model vulnerability relationships over dependency structure with the goal to enrich SBOMs.

This is different to the existing SBOM tool outputs where they are treated only as independent records. They visualize them as heterogeneous graphs where nodes represent the SBOM components and dependencies, the known software vulnerabilities, and the known software security weaknesses. The goal of their novel approach is to predict cascaded vulnerabilities in SSCs from SBOMs.

Research conducted in [24] claim that two key gaps exist in software supply chain security in general: the lack of accurate whole-ecosystem vulnerability propagation analysis and the absence of quantitative indicators for assessing vulnerability propagation impact. They address these limitations by proposing a vulnerability propagation analysis and a respective scoring system. Their work differ from our work as they do not address SBOM tools but try to generally answer the question on how widely and deeply vulnerability propagates. Their findings are complementary with our findings in the sense that there is a lack of accurate and complete vulnerability propagation analysis as such. Tackling the problem of currently little integration between SBOMs and Software Composition Analysis (SCA) tools the authors [25] propose a novel knowledge graph-based methodology for integrating vulnerability and dependency data into a holistic view on both aspects. With this contribution of a novel graph-theoretic methodology they improve visibility into how vulnerabilities propagate through complex dependencies. However, as they claim themselves they do not account for whether the vulnerable code is actually invoked or not. Meaning, the initial problem of obtaining a high amount of false positive information is not resolved.

Along with that, the study [26] investigates the contagion mechanisms of information security risks in software supply chains with the goal to identify key factors influencing risk propagation. The authors claim that no studies have been specifically conducted to elaborate a deeper understanding on vulnerability source, propagation, localization, and repair in SSCs. Their finding aligns with our finding during our literature review. Their results show that the risk transmission rate of software supply chain information security is influenced by the attack path. Also, they demonstrate the cascading effects of security risks in such multi-layer supply chain networks and by doing so, offer actionable insights for the design of predictive risk assessment in complex software supply chain ecosystems. They argue that this is especially relevant in the context of SSCs because – and we have also started our paper introduction with this key fact - they are multi-level structures and have complex interdependencies which has introduced new challenges in making them secure. Their argument support our finding on the fact that understanding propagation is currently missing and needed to propose effective SSC defence strategies. Similar conceptual approach is proposed in [27] where they address risk measures based on contagion paths and analyse how certain propagation dynamics influence optimal defence strategies. Further research [28] state the relevancy of understanding propagation effects because the propagation effect transforms the cyber risk into a systemic risk, which affects the whole supply chain ecosystem simultaneously. This makes it practically uncontrollable and consequently all available defence mechanism useless from a certain point on. As a result of their study the authors show that higher cyber risk is significantly associated with increased systemic risk. As mentioned in our introduction, propagation effects can transform a cyber risk into a systemic risk affecting whole ecosystems beyond the initial attack surface.

The paper [29] addresses a critical gap in traditional risk assessment methods, which primarily focus on individual risks or pairwise interactions (between two risks). The authors argue that these methods are insufficient because, in reality, risks often interact in higher-order structures (groups of three or more), leading to complex propagation effects that are not captured by simple pairwise models. The main objective is to develop and validate a novel risk assessment framework that can effectively model these higher-order interactions and their dynamic propagation through a network. This finding adds another element to our position on how important it is to include propagation effects when thinking about securing a SSC.

The findings of these mentioned related work emerge around propagation effects as the main conclusion that capturing and integrating these effects are currently missing when it comes to effective strategies in securing SSCs. While the main contributions of related work concentrates around ideas ranging from increasing the visibility aspects of SBOM tools, analysing the attack paths to think about defence strategies, addressing the (in)effectivity and (in)efficiency of SBOM tools in practices we follow a different approach. We will demonstrate our approach in the next section with a concrete experiment.

# 3 Methodology

## 3.1 Four-Stages Propagation Model

Our paper tries to find an answer to the introduced problem statement of SBOM tools specifically addressing the missing detection possibility of information that are relevant to depict propagation effects in a SSC. As part of our proposed solution to the problem we introduce a dynamic propagation model consisting of four stages. With this model we can very concretely demonstrate where the limitations of existing SBOM tools are. This, as we apply a very structured way with the four-stages model that represents the cascade path of a vulnerability exploitation. Our experiment results also give a first practical suggestion on how to potentially re-design the concept of SBOM tools to enhance SSC security. The experiment outlined in detail below will be performed by running four SBOM tools through all stages of our Propagation Model as following:

**Table 1.** Proposed Propagation Model Stages for the experiment Log4j

| Stage | Name | Description |
|---|---|---|
| **1** | **Structural Exposure** | The vulnerable component (log4j-core 2.14.1) is present in the software — either as a direct or transitive dependency. The attack surface exists but has not yet been activated. |
| **2** | **Vulnerable Class Presence** | The specific vulnerable class (JndiLookup) is physically present on the application classpath and is not disabled or removed. Presence does not confirm invocation. |
| **3** | **Code Reachability** | The vulnerable JndiLookup.lookup() method is reachable from the application's call graph — i.e., there exists at least one execution path from the application code to the vulnerable function. |
| **4** | **Taint Path Confirmation** | User-controlled input (e.g., HTTP headers, query parameters, form fields) flows into a Log4j logging call, providing the attacker with a vector to inject a malicious JNDI expression. |

### 3.2 Hypothesis to test

The main goal of this experiment is to analyse for a real-world affected software project the following:

- Can the chosen (open-source) SBOM tools detect the presence of the Log4Shell vulnerability (CVE-2021-44228); and
- Are they capable of assessing its exploitability and potential for dynamic propagation?

More detailed, the hypothesis we are testing with the experiment are as following:

- **Hypothesis 1** [2]: SBOM tools cannot distinguish between a) an application that logs HTTP headers through Log4j (fully exploitable) and b) an application that uses Log4j only for internal debug messages with no user-controlled input (not exploitable in practice): Both will receive the same CVE-2021-44228 CRITICAL flag. Regardless of whether the vulnerable JndiLookup class is ever invoked.
- **Hypothesis 2:** None of the evaluated open-source SBOM tools are capable of detecting dynamic propagation as defined. Hence SBOM tools do not provide information to assess if the successful exploitation of CVE-2021-44228 would enable lateral movement and cascading compromise of further systems.

### 3.3 Experiment Design: Tools and Projects chosen

In this section the experimental design consisting of SBOM tools under review, projects selection and the chosen vulnerability is outlined.
**Selection of SBOM Tools.** As outlined in the section of *Related Work* several research papers have already analysed the existing and most prominently used SBOM tools. We rely on these findings and select for experiment four of the most used open source SBOM tools as mentioned in the literature review (reference is made specifically to [20]). These are Syft, Trivy, Grype, and CycloneDX Generator (cdxgen). Syft generates SBOMs, Grype scans those SBOMs for vulnerabilities and Trivy serves as an all-in-one scanner for vulnerabilities.

**Table 2.** Overview on SBOM tools selected for the experiment

| Tool | Role |
|---|---|
| Syft (Anchore) | SBOM Generation |
| Trivy (Aqua Security | SBOM Generation + Vulnerability Scan |
| Grype (Anchore) | Vulnerability Scanner (consumes Syft SBOM) |
| cdxgen (CycloneDX/OWASP) | SBOM Generation (build-time) |

In addition to these SBOM tools we also used CodeQl and Semgrep for our experiment. The rational will be explained further below.

### 3.4 Rational of choosing Log4j's Log4Shell vulnerability

Researches and industry agree what was specified in [22] that Log4Shell (CVE-2021-44228) has been referred to as one of the most significant cyber security vulnerabilities in the modern age. Its Common Vulnerability Scoring System (CVSS) rating is 10 being the highest rating possible. The significance of the attack is also underscored because of Log4j's prevalence in numerous systems. This widely-used Java logging library propagated structurally into an estimated [30] 17,000+ Java packages on Maven Central alone, and then propagated as an exploit across hundreds of millions of devices globally within hours of disclosure. This propagation effect makes it very suitable for our experiment. Also the vulnerability posed a new challenge for securing the SSCs not only during the outbreak but also long after. While a response to Log4Shell was initially difficult as visibility was missing on the affected third-party components there is also a documented longitude impact proving that exploitation of the vulnerability continues even today. The other relevant aspect to choose it for our experiment is the availability of open source data for several projects that have been affected by Log4shell. This makes our experiment practically reproducible.

## 4 Experimental Setup

### 4.1 Selected Projects

Three Java-based projects were selected: Apache Solr 8.11.0, Spring Boot 2.5.0, and Ghidra 10.0.4. They were chosen because they represent different dependency and packaging models and each included a vulnerable Log4j component.

**Table 3.** Overview on Selected Projects and Log4j version

| Project | Ecosystem Ground Truth | Version | Log4j version |
|---|---|---|---|
| Apache Solr 8.11.0 | Maven | org.apache.solr:solr-core:8.11.0 | log4j-core:2.14.1 |
| Spring Boot 2.5.0 | Gradle runtimeClasspath | Spring Boot log4j smoke test module | log4j-core:2.14.1 |
| Ghidra 10.0.4 | JAR files | packaged distribution case | log4j-core:2.12.1 |

Together, the projects allowed comparison of SBOM tool behaviour across Maven metadata, Gradle project context, resolved JARs and packaged application distributions.The selected project have been run through our proposed four-stages propagation model.

### 4.2 Stage 1 Test: Structural Exposure

Stage 1 checks whether the vulnerable log4j-core component is present in the analysed project. At this stage, the goal is only to confirm component-level exposure and no claim is made about class usage, reachability, or exploitability.
**Result.** At this stage, the SBOM tools were generally effective because they are designed to identify software components. Syft detected log4j-core reliably when it was given a suitable input. Trivy SBOM mode detected the component in some input modes but missed it in others. Cdxgen also detected the component, but required the correct project or artifact context, such as Maven metadata, the full Gradle project structure, or deep JAR analysis. Grype and Trivy vulnerability scanning were treated as supplementary evidence rather than primary component-discovery tools. They confirmed Log4j CVEs only when the analysed SBOM that already contained the log4j-core component.
**Interpretation.** Stage 1 was confirmed for all three projects. The results show that SBOM tools are useful for structural exposure detection, but their results depend strongly on the input method. A missed result in one SBOM input mode does not necessarily mean the component is absent but it may mean the tool did not fully interpret that input format.

### 4.3 Stage 2 Test: Vulnerable Class Presence

Stage 2 assessed whether the specific vulnerable class was physically present inside the detected log4j core artifact: org/apache/logging/log4j/core/JndiLookup.class.
This stage is more specific than Stage 1. Detecting log4j core is not enough because Stage 2 requires proof that the vulnerable class itself exists inside the analysed JAR and has not be removed.
**Result.** For all three projects, direct JAR inspection was the primary evidence for Stage 2. This was necessary because standard SBOM outputs usually describe software at the package or component level. They can identify log4j-core, but they do not always enumerate every internal Java class file. The SBOM tools therefore had different usefulness at this stage. Syft provided class-level evidence only when the log4j JAR was extracted and scanned with file metadata enabled. In normal CycloneDX output, Syft mainly confirmed the parent component. Cdxgen provided class-level support for Ghidra through deep JAR analysis, where it reported JndiLookup as namespace/class-level evidence. Trivy SBOM mode, Grype, and Trivy vulnerability scanning did not provide reliable physical proof that JndiLookup.class was present.
**Interpretation.** Stage 2 was confirmed for all three projects primarily through direct JAR inspection. SBOM tools were useful as supporting evidence in selected cases but normal SBOM output should not be treated as sufficient proof of internal class presence. This stage confirms that the vulnerable class exists but it still does not prove invocation, reachability, or exploitability.

### 4.4 Stage 3 Test: Reachability Analysis

Stage 3 asks whether the vulnerable method is reachable from this application code: org.apache.logging.log4j.core.lookup.JndiLookup.lookup(). It requires stronger evidence: a call path from the application to JndiLookup.lookup().

The kind of source code pattern that would support Stage 3 would look like this: JndiLookup lookup=new JndiLokup(); lookup.lookup(...) or any application call that CodeQl can resolve to: org.apache.logging.log4j.core.lookup.JndiLookup.lookup(). To answer that, a tool must build or approximate a call graph. It must understand method calls, caller-callee relationships, inheritance, interfaces, dynamic dispatch, framework behaviour and sometimes reflection or runtime configuration.
**Result.** The SBOM tools were checked to see whether they could provide reachability evidence. Syft, Trivy SBOM mode, and cdxgen produced component or class-inventory evidence, but they did not provide a method-level call graph. Grype and Trivy vulnerability scanning also did not prove reachability. They correctly matched vulnerable Log4j packages to CVEs, but those matches were based on package metadata.
**Interpretation.** Why didn't SBOM tools detect Stage 3 reachability? The SBOM tools did not detect JndiLookup.lookup() reachability because reachability is not the same as component or class presence. SBOM tools such as Syft, Trivy SBOM mode and cdxgen are mainly designed to describe software inventory. Therefore, the absence of Stage 3 evidence in SBOM tools does not mean the application is safe and it does not mean reachability is impossible. It means that these tools are not designed to prove method-level reachability. SBOM output does not contain this kind of execution-path information. In addition, Grype and Trivy vulnerability scanning did not detect Stage 3 because they match vulnerable package versions to CVEs.

As we have demonstrated the limitations of SBOM tools for Stage 3, we used dedicated program-analysis tools such as CodeQL, bytecode call-graph analysis, or reachability-aware SCA tools to perform the experiment.
**Test with CodeQL analysis.** CodeQL was used as a stronger source-code analysis method. The CodeQL analysis asked two direct questions:

- Does the application directly call JndiLookup.lookup()?
- Does the application directly reference the JndiLookup type?

For Apache Solr 8.11.0, Spring Boot 2.5.0, and Ghidra 10.0.4, the same Stage 3 interpretation was applied. CodeQL was used to search for direct calls to org.apache.logging.log4j.core.lookup.JndiLookup.lookup() and for direct source-level references to the JndiLookup type. In all three analysed projects, both queries returned only CSV headers ("call","col1") which means that no direct call to JndiLookup.lookup() and no direct reference to the JndiLookup class were found in the analysed source code. For Ghidra 10.0.4, the cdxgen reachability-style search also returned zero reachability matches. Therefore, although dependency or JAR/class inspection may show that the vulnerable Log4j component is present, this is treated only as prerequisite evidence and not as proof of actual reachability.

**Interpretation.** Stage 3 was not confirmed for Apache Solr, Spring Boot, or Ghidra. The important conclusion is that SBOM tools and vulnerability scanners can identify vulnerable components and CVEs, but they do not prove method reachability. It only means that the vulnerable package version is present.

### 4.5 Stage 4 Test: Taint Path Analysis

The main question for this stage is: Does user-controlled input reach a logging sink? Stage 4 evaluates whether user-controlled input can flow into a logging call. In the context of Log4Shell, this is important because exploitation requires attacker-controlled data, such as an HTTP header, query parameter, request URI, or form value to reach a logging sink. If attacker-controlled input is logged and the runtime logging backend processes log4j lookups, then the application may have a possible exploitation path.

**Result.** The SBOM tools were checked for taint-flow evidence. Syft, Trivy SBOM mode and cdxgen did not provide source-to-sink data-flow information. They generated component, file, or class inventory evidence but they did not show whether HTTP headers, query parameters, request bodies, form fields, or other external inputs flow into logging calls. Grype and Trivy vulnerability scanning also did not provide Stage 4 evidence. These tools reported vulnerability metadata for vulnerable Log4j versions but they did not show a user-input-to-logger path.

**Interpretation:** Why didn't SBOM tools confirm Stage 4? SBOM tools cannot confirm Stage 4 because taint flow is a program-analysis problem not an inventory problem. A normal SBOM can list components, versions, package URLs, hashes, files and dependencies. It does not normally model HTTP request sources, query parameters, headers, request bodies, logger method calls.

As we have demonstrated the limitations of SBOM tools for Stage 4, that kind of evidence requires a taint analysis tool such as Semgrep taint mode.

**Test with Semgrep analysis.** The rules treated request-controlled or externally influenced values as sources such as: getHeader(...), getParameter(...), getRequestURL(...), and @RequestBody while logging calls such as log.debug(...), log.warn(...), Msg.error(...), and Msg.info(...) were treated as sinks. The goal was not to prove exploitation, but to identify candidate paths where potentially untrusted input could reach logging APIs.

*Result Apache Solr 8.11.0.* Semgrep identified a manually reviewed candidate flow where servletReq.getHeader("Authorization")reached log.debug(...).
This is considered a candidate path because the value originates from an HTTP request header which is normally supplied by the client and may therefore be attacker-controlled. If this value is passed into a logging call without being fully controlled or sanitized, attacker-provided text can enter the logging system.
*Interpretation.* In a Log4Shell-style analysis this is security-relevant because the risk depends not only on the presence of vulnerable Log4j components, but also on whether untrusted input can reach a logger. However, this remains candidate evidence rather than confirmed exploitation, because additional conditions must still be satisfied: the code path must execute, debug logging must be enabled, the value must remain attacker-controlled, the logging backend must route to vulnerable Log4j Core, lookup processing must be active and Stage 3 reachability to JndiLookup.lookup() must be confirmed.
*Result Ghidra 10.0.4.* Semgrep identified a weaker candidate flow where testScriptFile.getAbsolutePath() is concatenated into a message passed to Msg.error(...). This is considered a candidate path because a variable value, rather than a fixed constant, reaches a logging API.
*Interpretation.* However, the source is weaker than in Solr: a file path may be influenced by a user-selected script, project file, or local application state, but the report does not prove that it is directly attacker-controlled. Therefore, this finding requires further manual validation before it can be treated as a realistic user-input-to-logger path.
*Result Spring Boot 2.5.0.* Semgrep returned zero taint findings. The only identified logging call was logger.info("Hello World") which logs a constant string. Since no external or user-controlled value reaches the logger, no candidate Stage 4 path was found.

For reproducibility, all Stages (1-4) analysis code is available in the accompanying GitHub repository:
https://anonymous.4open.science/r/log4j-propagation-analysis-CBED.

## 5 Discussion and Conclusions

**A Propagation-Centred approach.** The novelty of our approach compared to other scientific work is that we start our analysis from the perspective of propagation occurrence. When we have knowledge of it i.e. at which stage the vulnerability currently is in the supply chain, the exploitability risk becomes controllable. We introduced a four-stages propagation model and based on that investigate what is technically needed to obtain information of every single stage of the model with the goal to obtain knowledge about where on the propagation path the vulnerability currently finds itself. Our central contribution is placing propagation effects of cyber-attacks in SSCs at the core of our research. As our experiment with SBOM tools makes it concrete, existing tools stop at structural dependency discovery and CVE matching. They do not systematically assess reachability, call-graph exposure, or cascading compromise. This gap is precisely what our propagation model addresses. Understanding and obtaining information on propagation effects is ultimately essential for increasing SSC security: it makes SSC risk controllable, provides information for differentiated defence strategies, and can prevent a cyber risk from developing into a systemic risk affecting the whole organisation and its (eco)system.

**Conclusion of our experiment.** As demonstrated concretely by our experiment existing SBOM tools primarily support structural dependency discovery and CVE matching. We conclude that this limited information corresponds to the gap between Stage 2 and Stage 3 of our propagation model. This gap directly explains the well-known problem of SBOM tools flagging vulnerabilities in unreachable code and producing systematically high false-positive rates and not assessing whether a vulnerable component is present at class level, reachable in the application call graph and exposed to attacker-controlled input. Our propagation model makes these results theoretically grounded rather than merely empirical: the missing information has a structural cause. At the same time, we have shown that certain tools used in practice do provide propagation-relevant information reflected in Stage 3 and Stage 4 of our model, confirming that closing this gap is technically feasible.

**Future Work.** Our experiment and applied propagation model demonstrate that the gap between Stage 2 and Stage 3 is the critical missing layer in current SBOM tool design. This has a direct consequences for SSC cyber risk controllability. Also we have analysed one example of a critical finding for Apache Solr 8.11.0. and by doing so we demonstrate the practical potential of the propagation information obtained with our experiment for future work. It goes beyond the requirement of just enhancing the visibility features. Consequently, we suggest further research into how SBOM tools can be enhanced to systematically capture propagation-stage information. This framing transforms an empirical observation into a theoretically grounded finding with clear implications for future SBOM tool design. We will continue to extend our propagation model by adding further stages, with the aim of understanding how propagation dynamics affect SSCs more broadly. Also, we aim to extend our findings into generalized results to support applicability for different environments and network-alike structures beyond SSCs.

**Disclosure of Interests.** The authors have no competing interests to declare that are relevant to the content of this article.